\documentstyle[12pt,a4,epsf]{article}
\begin{document}
\input psfig
\bibliographystyle{unsrt}
\baselineskip= 18pt
\pagenumbering{arabic}
\pagestyle{plain}
\def\bit{\begin{itemize}}
\def\eit{\end{itemize}}
\def\half{{1\over 2}}
\def\OO{\Omega}
 \def\aa{\alpha}
 \def\bb{\beta}
 \def\bk{{\bf k}}
 \def\bkp{{\bf k'}}
 \def\bqp{{\bf q'}}
 \def\bq {{\bf q}}
 \def\EE{\Bbb E}
 \def\EEx{\Bbb E^x}
 \def\EEo{\Bbb E^0}
 \def\LL{\Lambda}
 \def\PP{\Bbb P^o}
 \def\rr{\rho}
 \def\SS{\Sigma}
 \def\ss{\sigma}
 \def\ll{\lambda}
 \def\dd{\delta}
 \def\ww{\omega}
 \def\ll{\lambda}
 \def\DD{\Delta}
 \def\DDt{\tilde {\Delta}}
 \def\kr{\kappa\lb \LL\rb}
 \def\PPx{\Bbb P^{x}}
 \def\gg{\gamma}
 \def\kk{\kappa}
 \def\tt{\theta}
 \def\eps{\epsilon}
 \def\lb{\left(}

 \def\rb{\right)}
 \def\prt{\tilde p}
\def\pt{\tilde {\phi}}
 \def\hal{{1\over 2}\nabla ^2}
 \def\bg{{\bf g}}
 \def\bx{{\bf x}}
 \def\bu{{\bf u}}
 \def\bv{{\bf v}}
 \def\by{{\bf y}}
 \def\hag{{1\over 2}\nabla}
 \def\beq{\begin{equation}}
 \def\eeq{\end{equation}}
 \def\bea{\begin{array}}
 \def\eea{\end{array}}
 \def\cosech{\hbox{cosech}}
 \def\fr{\frac}
 \def\Tr{{\mbox{Tr}}}

\title{Mixing Scenarios for Lattice String Breaking.}
\author{I T Drummond  \\
        Department of Applied Mathematics and Theoretical Physics \\
        University of Cambridge \\
        Silver St \\
        Cambridge, England CB3 9EW}
\maketitle
\begin{abstract}
We present some simple scenarios for string breaking on the lattice
based on a crude strong coupling model introduced previously.
We review the dependence of the model on lattice spacing and extend it to include
degenerate dynamical quarks and also meson exchange diagrams. A comparison is 
made between quenched and unquenched calculations. We examine string
breaking in the presence of a static quark-diquark system, a situation
that is specific to $SU(3)$~.
\end{abstract}
\vfill
DAMTP-98-103
\pagebreak

\section{Introduction}
String breaking has importance in lattice calculations because it reveals 
directly the effect of dynamical quarks. However the energy crossover 
effect characteristic of string breaking has proved difficult to observe 
\cite{DQ1}-\cite{DQ7}.  Recently progress has been made and string breaking 
has been observed as a mixing phenomenon directly in $SU(2)$ Higgs models in 
three and four dimensions \cite{PhWi,KnSo,DeT}. The phenomenon is currently 
under investigation in the much more demanding case of QCD \cite{Burk,Ka}.
A crude model of the process based on strong coupling ideas has
been proposed which pictures it as a mixing process between a string
state and a two-meson state \cite{ITD1,ITD2}. It shows that because of its 
dependence on a mixing angle the Wilson loop may not reveal the string breaking
energy crossover unless the mixing region is sufficiently broad and that in any case
it is not by itself a satisfactory observable for revealing the crossover. This is consistent
with numerical simulations \cite{PhWi,KnSo} which show the need for a 
suite of operators with members appropriate to both the string and two-meson states.

In this paper we develop the crude strong coupling model to
discuss a number of simple mixing scenarios and use it to 
discuss the relationship between quenched and unquenched calculations.
We take the opportunity to improve the model by including energy factors
associated with the emission of light quarks from bound states.
These factors are required in order to achieve results independent of the 
lattice spacing. They were missed in the original presentation \cite{ITD1} but are 
essential parameters \cite{ITD2,PW}. 

We elaborate the model on the one hand by including
several quark flavours and on the other by including
diagrams that represent meson exchange.

Finally we apply our approach to more complex 
situations that are specific to $SU(3)$ and examine the case of
string breaking in the presence of a static quark-diquark system.

\section{Rules for Simple Planar Model}

The model we use was explained previously \cite{ITD1,ITD2}. It is based on simple 
planar lattice diagrams with internal dynamical quark loops. The initial
formulation of the model is in terms of the rules for a 
``strong coupling expansion'' \cite{CREUTZ, MonMun}, modified to accommodate
the presence of bound states in a simple way. The rules are (slightly modified
from those of ref \cite{ITD1} as indicated in ref \cite{ITD2}):
\bit
\item[1.] A factor of $e^{-\ss}$ for each plaquette, where $\ss$ is
the (dimensionless) string tension.  
\item[2.] A factor of
$$
2\kk\left(\fr{1+\gg.e}{2}\right)
$$
for each Wilson quark line in the direction of the unit vector $e$,
interior to the diagram. Here $\kk$ is the standard quark hopping parameter.
\item[3.] A factor of
$$
2\kk'\left(\fr{1+\gg.e}{2}\right)
$$
for each Wilson quark line propagating in the direction of the unit vector $e$
along a static anti-quark line. This creates the model for the static-light
bound state meson.
\item[4.] A factor of $\sqrt{W}$ for each emission of a light quark from a 
meson state. Here $W$ is a (dimensionless) energy parameter
that determines the rate of emission of light quarks from the meson.
Such a parameter is essential in order that the results of the model are
independent of the lattice spacing.
\item[5.] A factor of $(-1/3)$ (in the case of $SU(3)$) for each internal quark loop.
\item[6.] A trace over the spin matrix factors for each internal quark loop.
\eit

In the above the hopping parameter, $\kk$, reflects the light quark mass, $2\kk=e^{-m_q}$,
and the bound state hopping parameter, $\kk'$, determines the energy of the
static meson, $2\kk'=e^{-E_M}$~.
(In ref \cite{ITD1} $E_M$ was the energy of a two-meson state and therefore had
twice the value assigned here.)

\begin{figure}[htb]
\begin{center}\leavevmode
\epsfxsize=10 truecm\epsfbox{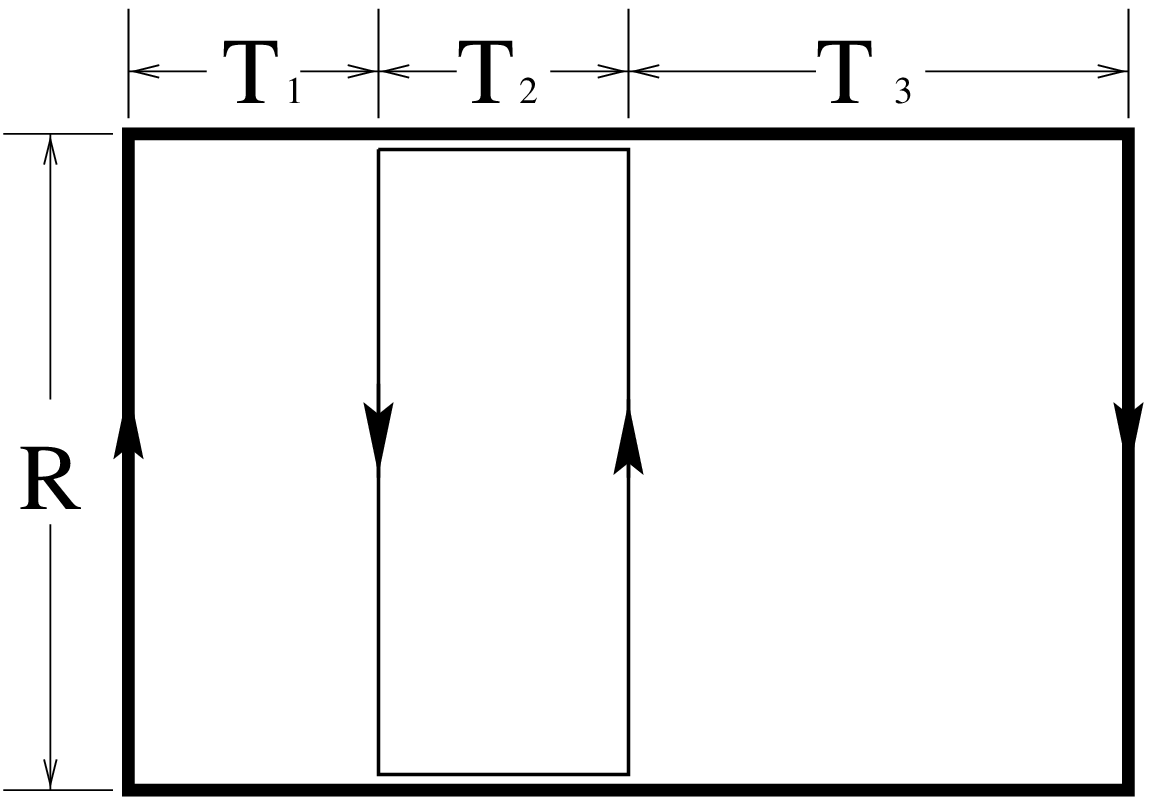}
\end{center}
\caption[]{Wilson loop (heavy line) containing internal quark loop (light line).}
\label{figure:F1}
\end{figure}

If we apply the above rules to the diagram in Fig. \ref{figure:F1} we obtain for
the string-string transition amplitude the result
\beq
{\cal G}_{SS}=\left(e^{-\ss R}\right)^{T_1}
\fr{1}{\sqrt{6}}We^{-m_qR}\left(e^{-2E_M}\right)^{T_2}
                  \fr{1}{\sqrt{6}}We^{-m_qR}\left(e^{-\ss R}\right)^{T_3}~~.
\eeq

\section{Lattice Spacing Dependence}

The string breaking scenario considered previously involved
a static quark and anti-quark separated by a (dimensionless) distance $R$.
In the notation of previous papers \cite{ITD1,ITD2}, we have
\beq
a=e^{-\ss R}~~,~~b=e^{-2E_M}~~,~~c=\fr{1}{\sqrt{6}}We^{-m_qR}~~.
\eeq
The basic string state has an energy $\ss R$ and the two-meson state 
an energy $2E_M$. As shown in ref \cite{ITD1} the propagating eigenstates 
have energies $E_\pm=-\log\ll_\pm$
where
\beq
\ll_\pm=\fr{1}{2}\left\{a+b\pm\sqrt{(a-b)^2+4abc^2}\right\}~~.
\eeq
To pick out the lattice spacing independent part we can treat all
the dimensionless parameters as $O(a_L)$, where $a_L$ is the lattice spacing. 
We then treat $R$ and any time interval, $T$, involved as $O(a_L^{-1})$ 
and expand to the lowest significant order in $a_L$.  We then find
\beq
E_\pm=\fr{1}{2}\left\{\ss R+2E_M\pm\sqrt{(\ss R-2E_M)^2+\fr{2}{3}W^2e^{-2m_qR}}\right\}~~.
\eeq
Having carried out this operation we can treat the above
formula as referring to dimensionfull quantities. Note that it is essential to
have the energy parameter $W$ as a factor in the quantity $c$ in order that this result
holds true \cite{ITD2,PW}.

Similar remarks apply to other quantities in the model. The mixing angle is
\beq
\tan\tt=\fr{-(a-b)+\sqrt{(a-b)^2+4abc^2}}{2\sqrt{ab}c}~~.
\eeq
Carrying out the expansion in small quantities we find
\beq
\tan\tt=\fr{\ss R-2E_M+\sqrt{(\ss R-2E_M)^2+\fr{2}{3}W^2e^{-2m_qR}}}
                                                 {\sqrt{\fr{2}{3}}We^{-m_qR}}~~.
\eeq

The formulae in ref \cite{ITD1} for the mixing range $\Delta R$ and the
energy split at maximal mixing, $\Delta E$ are correct when the energy parameter 
$W$ is included. As indicated in ref \cite{ITD2}, the results are
\beq
\Delta R=\pi\sqrt{\fr{2}{3}}\fr{W}{\ss}e^{-m_qR}~~,~~
\Delta E=\sqrt{\fr{2}{3}}We^{-m_qR}~~.
\eeq
The relation 
\beq
\Delta R=\pi\fr{\Delta E}{\ss} 
\label{DRDE}
\eeq
is preserved. All these formulae may 
now be regarded as being explressed in physical parameters.  
The transition energy parameter is set by $\Delta E$. Eq (\ref{DRDE}) is then a 
prediction for the mixing range in terms of the maximal energy split.

\section{Quenched/Unquenched Comparison}

An important reason for measuring string breaking is to identify the effect of
dynamical fermions directly. It is therefore useful to consider the 
comparison between relevant quantities in the quenched and unquenched situations.
A natural amplitude to measure is the transition amplitude between the string and
the two-meson state \cite{StKo}. The unquenched case was considered previously 
\cite{ITD1,ITD2}. The result was an amplitude of the form
\beq
{\cal G}_{MS}(T)=\sqrt{ab}\sin\theta\cos\theta(\ll_{+}^{(T-1)}-\ll_{-}^{(T-1)})~~,
\eeq
where $T$ is the full interval in the time direction associated with the measurement.
The significance of this form is that the amplitude is suppressed outside the
mixing region where the factor $\sin\theta\cos\theta$ vanishes. If we
reexpress the formula in terms of physical parameters then we find
\beq
{\cal G}_{MS}(T)=\fr{1}{2}\left(\fr{\sqrt{\fr{2}{3}}We^{-m_qR}}
{\sqrt{(\ss R-2E_M)^2+\fr{2}{3}W^2e^{-2m_qR}}}\right)\left(e^{-E_{+}T}-e^{-E_{-}T}\right)~~.
\eeq

\begin{figure}[htb]
\begin{center}\leavevmode
\epsfxsize=10 truecm\epsfbox{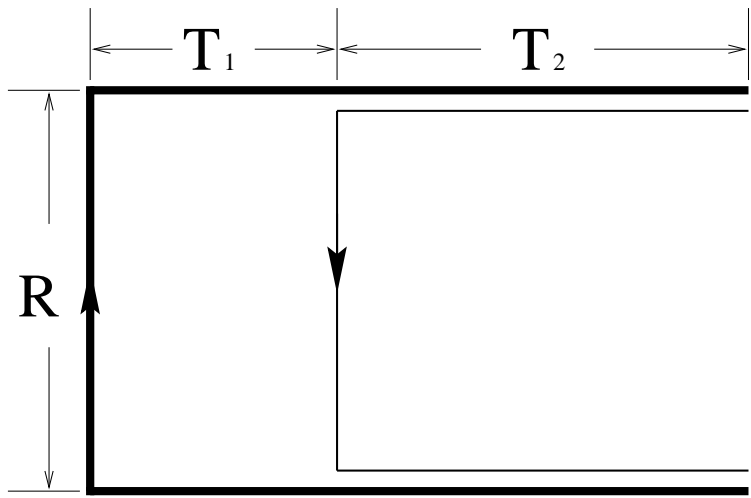}
\end{center}
\caption[]{Contribution to the transition amplitude in the quenched approximation.}
\label{figure:F2}
\end{figure}

The corresponding quenched approximation can be described in our model as a sum over
graphs of the form shown in Fig. \ref{figure:F2}. The resulting amplitude expressed in terms of
the parameters $a$, $b$ and $c$ is
\beq
{\cal G}^{(Q)}_{MS}(T)=\sum_{T_1=1}^{T-1}b^{T_2}ca^{T_1}~~,
\eeq 
where $T=T_1+T_2$~.The result is
\beq
{\cal G}^{(Q)}_{MS}(T)=cab\fr{b^{T-1}-a^{T-1}}{b-a}~~.
\eeq
When translated into physical parameters we find
\beq
{\cal G}^{(Q)}_{MS}(T)
=\fr{1}{2}\left(\fr{\sqrt{\fr{2}{3}}We^{-m_qR}}{\ss R-2E_M}\right)(e^{-2E_MT}-e^{-\ss RT})~~.
\eeq
The quenched and unquenched results are very similar in the region 
$|\ss R-2E_M|>\sqrt{\fr{2}{3}}We^{-m_qR}$~. In the complementary region, the inner
part of the mixing region on our definition, they exhibit very different behaviour.
The unquenched amplitude exhibits the continuous crossover while the quenched
amplitude shows an abrupt change from one exponential behaviour to the other
at the crossover point. It would be extremely interesting if this contrast
in behaviour could be verified in a real simulation. Although our model rather
oversimplifies the relationship of the quenched and unquenched calculations
it is still possible that a measurement of the above transition amplitude
in a quenched simulation would yield an estimate of the energy $W$ that
could stand as a prediction for the unquenched calculation.

\section{Several Flavours}

So far the model has been formulated only for a single flavour
of light quark. The presence of several flavours of quark is accounted
for by summing over the $N_f$ flavour insertions of quark loop.
This means that the model has $N_f+1$ channels,
namely the string channel and the $N_f$ two-meson channels associated 
with the different flavours of quark. It is sufficient to consider the case
$N_f=2$~. Following the same analysis as in the single flavour case
and assuming that the light quarks are degenerate we easily see that the 
summation of relevant graphs is achieved by considering a $3\times 3$ 
correlation matrix $G(T)$ that obeys the update equation
\beq
G(T+1)=AG(T)~~,
\eeq
where
\beq
A=\left(\bea{ccc}a&ac&ac\\
                bc&b&0\\
                bc&0&b\eea\right)~~,
\eeq
and we can assume that $G(T)=A^T$~.
It is immediately obvious that the two-meson state that is anti-symmetric
under the interchange of flavours propagates with the factor $b^T=e^{-2E_MT}$~.
It does not mix with the string state. The symmetric state does mix with the string.
If we choose the symmetric and antisymmetric states as basis states then the
matrix $A$ assumes the form
\beq
A=\left(\bea{ccc}a&\sqrt{2}ac&0\\
                \sqrt{2}bc&b&0\\
                0&0&b\eea\right)~~,
\eeq
The problem therefore remains a two-channel mixing problem of the same form as before
with the minor change that $c\rightarrow \sqrt{2}c$~. There is no need to 
carry the explicit calculation further since the results are obvious.

The general case of $N_f$ flavours follows the same lines. Only the completely
flavour symmetric combination of two-meson states mixes with the string.
The mixing being of a strength $\sqrt{N_f}$ times greater than the single flavour case.
The remaining $N_f-1$ two-meson channels orthogonal to the symmetric one remain unmixed.

If the light quarks are not flavour degenerate then to a first approximation
the two-meson state corresponding to the lightest quark will show the first and 
strongest mixing with the string state since it lies at the lowest energy.
Heavier quarks will mix at larger distances with decreasing strength.
It would be interesting to test this obvious and plausible scenario in a full
QCD calculation.

\section{Extended Planar Model}

\begin{figure}[htb]
\begin{center}\leavevmode
\epsfxsize=10 truecm\epsfbox{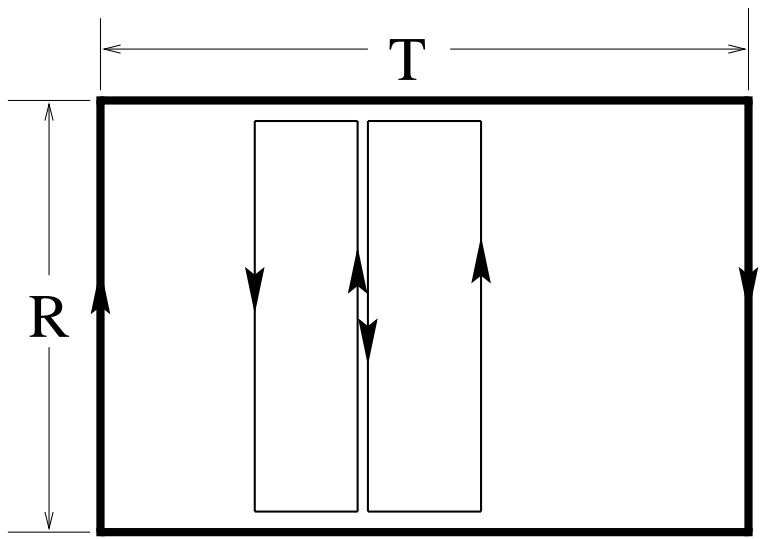}
\end{center}
\caption[]{Wilson loop contribution with a light quark-anti-quark exchange.}
\label{figure:F3}
\end{figure}

It was implicit in the summation technique used previously that diagrams of the form
shown in Fig. \ref{figure:F3} were not included \cite{ITD1,ITD2}. To include them 
requires a new rule for the vertex associated with the emission of the light
quark pair. It is

\bit
\item[7.] A factor $\sqrt{W_p}$ for the pair emission vertex, where
$W_p$ is a dimensionless energy parameter. Again this factor is necessary
in order to eliminate explicit dependence of the results on the lattice spacing.
\eit

If, after emission, such a light quark-anti-quark pair were to propagate
along a link, the corresponding factor according to the rules would be
\beq
\left(\fr{1+\gg.e}{2}\right)\otimes\left(\fr{1-\gg.e}{2}\right)(2\kk)^2
\eeq 
Naively the hopping parameter factor $(2\kk)^2$ is equal to $e^{-2m_q}$. 
In the context of the model it would not be unreasonable to identify the
quark pair exchange as a meson exchange. If the meson has a mass $m$ then 
we should replace $(2\kk)^2$ with $e^{-m}$~. This is the interpretation we will adopt.

For completeness we give the list of rules for computing 
diagrams that include strings, two-meson states and light meson exchanges.
\bit
\item[1.] A factor $a=e^{-\ss R} $ that
propagates the string by one time step.
\item[2.] A factor $b=e^{-2E_M}$ that
propagates the two-meson state by one time step.
\item[3.] A factor $c=(W/\sqrt{6})\left(2\kk\right)^R $
associated with the transition from string to two-meson state
and vice-versa.
\item[4.] A factor $\delta=\fr{1}{6}W_p e^{-mR}$ for each quark-anti-quark exchange.
\eit
As in the original model the summation can be carried out by computing a $2\times 2$
matrix of propagators connecting the string and two-meson states
\beq
G(T)=\left(\begin{array}{cc}G_{SS}(T)&G_{SM}(T)\\ G_{MS}(T)&G_{MM}(T)\end{array}\right)~,
\eeq
which obeys the update equation
\beq
G(T+1)=AG(T)~~,
\label{GF1}
\eeq
where
\beq
A=\left(\begin{array}{cc}a&ac\\ bc&b(1+\delta)\end{array}\right)~,
\eeq
The natural solution to eq(\ref{GF1}) is $G(T)=A^T$~.
As before $A$ can be expressed in the form $A=DO\Lambda O^{-1}D^{-1}$~,
where $D$ and $O$ have the forms
\beq
D=\left(\begin{array}{cc}\sqrt{a}&0\\0&\sqrt{b}\end{array}\right)~~,~~
O=\left(\begin{array}{cc}\cos\theta&-\sin\theta\\\sin\theta&\cos\theta\end{array}\right)~~,
\eeq
and $\Lambda$ is a diagonal matrix of eigenvalues,
\beq
\ll_{\pm}=\fr{1}{2}\left\{(a+b(1+\delta))\pm\sqrt{(a-b(1+\delta))^2+4abc^2}\right\}~~.
\eeq
The mixing angle $\theta$, that describes the overlap of the string and
two-meson channels with the eigenchannels of definite energy, $E_{\pm}=-\log\ll_{\pm}$~,
satisfies
\beq
\tan\theta=\fr{-(a-b(1+\delta))+\sqrt{(a-b(1+\delta))^2+4abc^2}}{2\sqrt{ab}~c}~~.
\eeq
The analysis of mixing is essentially the same as before except that the centre
of the mixing region where $\theta=\pi/4$ occurs when
$a=b(1+\delta)$ that is when $R=R_c$ where
\beq
\ss R_c=2E_M+\log(1+\delta)\simeq 2E_M+\fr{1}{6}W_p e^{-mR_c}~~.
\eeq
This is consistent with the idea that the extra light meson exchange represented by $\delta$ 
corresponds to a potential interaction between the two static mesons
of the form $(W_p /6)e^{-mR}$ that displaces the centre of the mixing region in
the appropriate way. If we estimate the mixing region as 
$\Delta R=\fr{\pi}{2}\fr{dR}{d\tt}|_{R=R_c}$ then we find
\beq
\Delta R=\pi\sqrt{\fr{2}{3}}\fr{We^{-m_qR_c}}{\ss-\fr{1}{6}mW_p e^{-mR_c}}~~.
\eeq
This is essentially the same result as derived previously \cite{ITD1,ITD2} but 
now including the effect of the potential between the mesons. As before
the energy split at the point of maximum mixing is $\Delta E=2c$, that is
\beq
\Delta E=\sqrt{\fr{2}{3}}We^{-m_qR}~~.
\eeq
We have therefore the relation between $\Delta R$ and $\Delta E$ of the form
\beq
\Delta R=\pi\fr{\Delta E}{\ss-\fr{1}{6}mW_p e^{-mR_c}}~~.
\eeq
These results may now be interpreted as referring to quantities in physical units.
In deriving them we have dropped corrections that vanish with the lattice spacing.

\section{Quark/Diquark String Breaking}

So far string formation and breaking in QCD has been sought in a
context in which the string joins a static quark and anti-quark pair.
An interesting alternative scenario involves a static-light diquark 
supporting a string along with another static quark. This is a situation that is 
particular to $SU(3)$ gauge theory. We again have quenched and  
unquenched possibilities.

\begin{figure}[htb]
\begin{center}\leavevmode
\epsfxsize=10 truecm\epsfbox{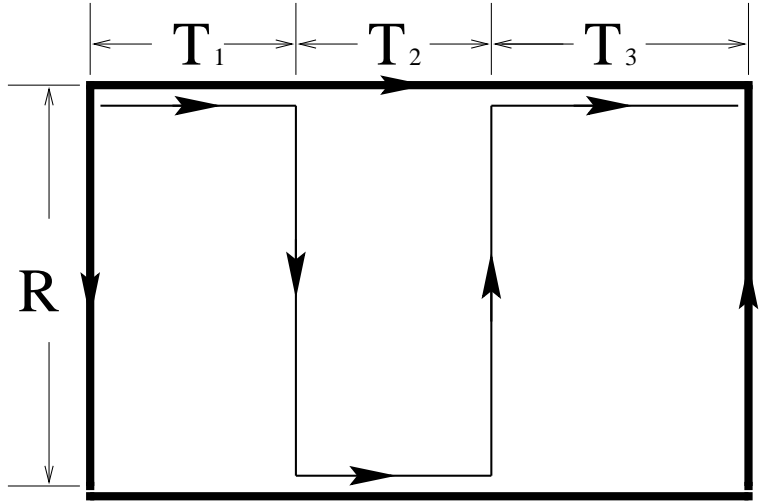}
\end{center}
\caption[]{Contribution to a heavy quark-diquark amplitude without internal 
quark loops. The path $C_1$ runs along the top of the graph. The path $C_2$
runs down the left side along the bottom and up the right side of the graph.}
\label{figure:F4}
\end{figure}

A relevant correlation function can be formulated in the following
way.
\beq
{\cal F}(T)=\langle\fr{1}{6}U(C_1)_{\aa_1\bb_1}U(C_2)_{\aa_2\bb_2}G_{q\aa_3\bb_3}(T)
               \epsilon_{\aa_1\aa_2\aa_3}\epsilon_{\bb\bb\bb}\rangle
\label{qdqcorr}
\eeq
Here the angle brackets indicate averaging over the gauge fields
and $G_{q\aa_3\bb_3}(T)$ is the light quark propagator from the origin 
to the origin over a time interval $T$~. The matrices $U(C_1)$ and $U(C_1)$
are each the product of gauge $SU(3)$ matrices along the paths
$C_1$ and $C_2$ that contain the static quarks and join the ends of the 
light quark propagator, as indicated in Fig. \ref{figure:F4}. 

In eq(\ref{qdqcorr}) we have suppressed the spin labels of the quarks.
However a feature of diquark dynamics is its sensitivity to the spins
of the constituent quarks \cite{HeKa}. For this reason it is important
in this case to take account of the static quark spins as well
as that of the light quark. We will do this by assuming that the 
static quark and light quark bind in a singlet state. The triplet
state, being higher in mass \cite{HeKa}, we will ignore for simplicity.
In that case lines in our diagramatic model that carry a diquark state
will have simple scalar propagators while the lines carrying a light or static
quark in the direction $e$ will have the standard spin factor $(1+\gg.e)/2$
as well as the appropriate energy exponential.

\begin{figure}[htb]
\begin{center}\leavevmode
\epsfxsize=10 truecm\epsfbox{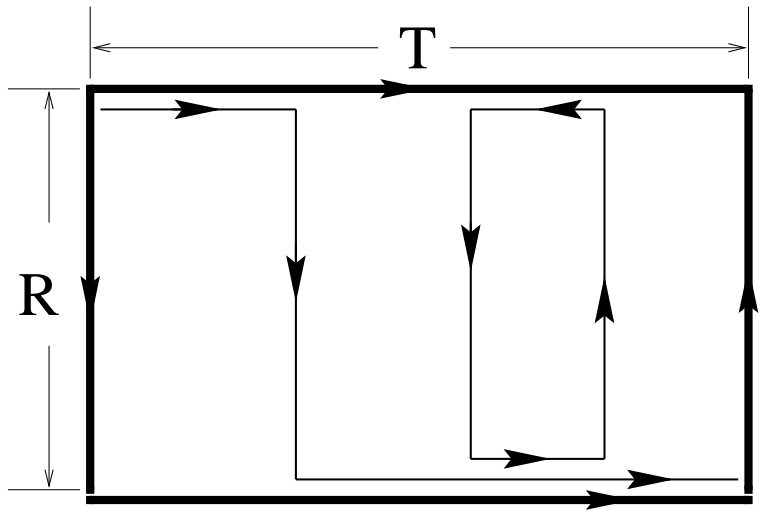}
\end{center}
\caption[]{Contribution to a heavy quark-diquark amplitude with an internal
quark loop.}
\label{figure:F5}
\end{figure}

A typical diagram that contributes to the amplitude in our simplified
planar diagram model is shown in Fig. \ref{figure:F4}. The light quark 
propagates along the static quark lines in a diquark combination and
also makes transitions between the static quarks on each side of the diagram.
This diagram is part of the quenched approximation.
The incorporation of dynamical quarks into the model results in 
diagrams such as that in Fig. \ref{figure:F5} where internal quark loops appear
creating intermediate states that involve a static nucleon, containing
one static and two light quarks, together with a static meson, containing a 
static quark and a light anti-quark.

\subsection{Quenched Calculation}

Because the light quark can make a transtion from one static quark to
the other, even the quenched calculation is not trivial. Within the model
we are concerned with two channels. The left channel in which the light quark 
propagates along the static quark at the origin and the right channel in which it
propagates along the static quark displaced a distance $R$ from the origin.
We will associate a hopping parameter $\kk''$ with the motion of the light quark along
a static quark line. We will introduce an energy $W_d$ that provides
a factor $\sqrt{W_d}$ for the strength of emission of the light quark from
a static diquark. With these rules the contribution to ${\cal F}(T)$ of the diagram 
in Fig. \ref{figure:F4} is
\beq
\left(2\kk''e^{-\ss R}\right)^{T_1}(\fr{W_d}{2}e^{-m_qR})\left(2\kk''e^{-\ss R}\right)^{T_2}
      (\fr{W_d}{2}e^{-m_qR})\left(2\kk''e^{-\ss R}\right)^{T_3}\left(\fr{1+\gg_0}{2}\right)
\eeq
The factors of $\fr{1}{2}$ that accompany the passage of the light quark across the diagram
follow from the $\gg$-matrix algebra. Because it plays no further r\^ole
from now on we will drop the quark spin factor in ${\cal F}(T)$~.
In order to sum up diagrams of the type we have just evaluated we require a $2\times 2$
matrix of propagators in order to describe the two channels in the problem. We have
\beq
G(T)=\left(\bea{cc}G_{LL}(T)&G_{LR}(T)\\G_{RL}(T)&G_{RR}(T)\eea\right)~~.
\eeq
Up tp terms that vanish with the lattice spacing, ${\cal F}(T)=G_{LL}(T)$~.
The summation over diagrams proceeds by requiring the propagators to obey the 
stepping relation
\beq
G(T+1)=A'G(T)~~,
\eeq
where
\beq
A'=\left(\bea{cc}a'&a'c'\\a'c'&a'\eea\right)~~,
\eeq
where $a'=2\kk''e^{-\ss R}=e^{-\ss R-E_d}$ and $c=\fr{1}{2}W_de^{-m_qR}$
and we have identified $E_d=-\log(2\kk'')$ as the static diquark energy.
It is immediately obvious that there are two eigenchannels. A symmetrical 
superposition of the left and right channels and an antisymmetric superposition. 

Choosing these as the basis channels $A'$ takes the form
\beq
A'= \left(\bea{cc}a'(1+c')&\\&a'(1-c')\eea\right)
\eeq
The eigenenergies of these channels respectively $\epsilon_{\pm}
=-\log(a'(1\pm c'))\simeq E_d+\ss R\pm\fr{1}{2}W_de^{-m_qR}$~.
We see here the influence of the quark exchange in splitting the degeneracy of the 
two string-diquark channels.

\subsection{Dynamical Quarks}

The effect of dynamical quarks is to insert light quark loops into the diagrams.
As pointed out above this creates the possibility of two new channels
one with a static nucleon at the origin and a static meson at a distance $R$
and the other with the nucleon and meson interchanged. See Fig \ref{figure:F5}~.

In order to sum over the diagrams of the model we need a $4\times 4$
matrix of correlators. 
\beq
G(T)=\left\{G_{ij}(T)\right\}~~,
\eeq
where $i,j=1,2,3,4$ and the labels 1 and 2 refer to the diquark-string
channels with the diquark on the left and right respectively ($L$ and $R$ above). 
The labels 3 and 4 refer to the nucleon-meson channels with the nucleon on the left 
and right respectively. The upgrade step is
\beq
G(T+1)=PG(T)~~,
\eeq
where now
\beq
P=\left(\bea{cccc}a'&a'c'&a'c''&0\\
                 a'c'&a'&0&a'c''\\
                 b'c''&0&b'&0\\
                 0&b'c''&0&b'\eea \right)~~,
\eeq
where $a'$ and $c'$ are as before and 
\beq
b'=e^{-(E_N+E_M)}~~,\mbox{and}~~c''=\fr{\sqrt{WW_N}}{\sqrt{6}}e^{-m_qR}~~.
\eeq
Here $W_N$ is the energy parameter associated wiith the emission of a light quark by
a static nucleon. The mixing phenomena in the model are
easily disentangled because the left-right symmetry of $P$ means that the symmetric
diquark-string state mixes only with the symmetric nucleon-meson state and similarly
for the corresponding antisymmetric states. If we choose these symmetric and anti-symmetric
states as the basis channels then $G(T)$ and $P$ have the form
\beq
G(T)=\left(\bea{cc}G^{(S)}(T)&0\\0&G^{(A)}(T)\eea\right)~~,~~
P=\left(\bea{cc}P^{(S)}&0\\0&P^{(A)}\eea\right)~~,
\eeq
and
\beq
P^{(S)}=\left(\bea{cc}a'(1+c')&a'c''\\b'c''&b'\eea\right)~~,~~
P^{(A)}=\left(\bea{cc}a'(1-c')&a'c''\\b'c''&b'\eea\right)~~.
\eeq
We have $G^{(S)}(T+1)=P^{(S)}G^{(S)}(T)$ and $G^{(A)}(T+1)=P^{(A)}G^{(A)}(T)$ with
the solutions $G^{(S)}(T)=\left(P^{(S)}\right)^T$ and $G^{(A)}(T)=\left(P^{(A)}\right)^T$~. 
We can write
\beq
P^{(S,A)}=DO^{(S,A)}\Lambda^{(S,A)}(O^{(S,A)})^{-1}D^{-1}~~,
\eeq
where 
\beq
D=\left(\bea{cc}\sqrt{a'}&0\\0&\sqrt{b'}\eea\right)~~,~~
O^{(S,A)}=
\left(\bea{cc}\cos\tt_{S,A}&-\sin\tt_{S,A}\\\sin\tt_{S,A}&\cos\tt_{S,A}\eea\right)~~.
\eeq
The eigenvalues of $P^{(S,A)}$ are the columns of $DO^{(S,A)}$ and the eigenvalues 
are the entries in the diagonal matrix
\beq
\Lambda^{(S,A)}=\left(\bea{cc}\ll^{(S,A)}_+&0\\0&\ll^{(S,A)}_-\eea\right)~~.
\eeq
It is easily established that
\beq
\ll^{(S)}_{\pm}=\fr{1}{2}\left\{a'(1+c')+b'\pm\sqrt{(a'(1+c')-b')^2+4a'b'c''^2}\right\}
\eeq
and
\beq
\ll^{(A)}_{\pm}=\fr{1}{2}\left\{a'(1-c')+b'\pm\sqrt{(a'(1-c')-b')^2+4a'b'c''^2}\right\}
\eeq
The mixing angles are given by
\beq
\tan\tt_S=\fr{-(a'(1+c')-b')+\sqrt{(a'(1+c')-b')^2+4a'b'c''^2}}{2\sqrt{a'b'}c''}~~,
\eeq
and a similar formula for $\tan\tt_A$ where $c'\rightarrow -c'$~.
The mixing analysis is very similar to that for the original string
breaking model and leads to the results that the critical value of the separation
in the symmetric and antisymmetric cases are given by $b'=a(1\pm c')$, that is
\beq
\ss R_c=E_N+E_M-E_d\pm\fr{1}{2}W_de^{-m_qR_c}~~,
\eeq
and the mixing ranges are
\beq
\Delta R_{S,A}=\pi\sqrt{\fr{2}{3}}\fr{\sqrt{WW_N}e^{-m_qR_c}}{\ss\pm\fr{1}{2}m_qW_de^{-m_qR_c}}~~,
\eeq
and the energy splits at maximum mixing are
\beq
\Delta E_{S,A}=\sqrt{\fr{2}{3}}\sqrt{WW_N}e^{-m_qR_c}~~.
\eeq
Although the energies are likelier to be higher than in the standard string
scenario the static quark-diquark string is also of great interest 
as an alternative string system specific to $SU(3)$ and because of what 
it can reveal about diquark dynamics.

\section{Conclusions}

We have explored further the consequences of a simple picture of
string breaking as a mixing phenomenon. We have clarified the
elimination of lattice spacing dependence in the model by improving it
over its original formulation with the introduction of appropriate
energy factors for quark and meson emission from bound states.
We have shown that it is possible to incorporate a meson-meson 
potential into the the original model and find, as might be expected, that it does not 
lead to radically different results on mixing.

We have also investigated the effect of degenerate light quark flavours.
Here the only significant effect is a slight strengthening of
the mixing effect by a factor $\sqrt{N_f}$~. It is also interesting
that only the totally symmetric combination of two-meson channels
participates in string breaking leaving the orthogonal combinations unmixed.

We point out that a comparison of the quenched and unquenched calculations
of string breaking is of considerable interest as a simulation. 
The appropriate measurement is of the string-meson transition amplitude.  
On the basis of the model we show that the two cases are likely to be rather
similar in the outer part of the mixing region but show distinct behaviours
in the central part of the mixing region, where the quenched amplitude
shows a sharp transition between energy exponentials and the unquenched case
shows a more gradual crossover. It is possible, if the comparison
between the two cases is made in an appropriate way
in an actual simulation, that the quenched measurement could provide a prediction
for the energy split and the size of the mixing region in the unquenched case.

We study static quark-diquark string breaking which shows some
simple but interesting patterns of channel mixing. The basic phenomenon
is very similar to that of the quark-anti-quark string breaking.
The symmetry of the system however provides us with two channels
in the quenched case. These each go on to mix with appropriately
symmetrical or antisymmetrical meson-nucleon channels. This pattern
of double string breaking is interesting in itself and more so
since it is specific to the case of $SU(3)$~. It seems worth
investigating this case even though the higher energies involved will make
it harder to measure. Perhaps an approach using a non-symmetric lattice
with a fine lattice spacing the time direction would be appropriate in this case.

The success of simulations using the $SU(2)$ Higgs model in observing
string breaking \cite{PhWi,KnSo} suggests
that similar results could be obtained for the scenarios discussed in this paper
with scalar matter fields replacing the dynamical quarks. It would of course
be necessary to go to the $SU(3)$ Higgs model to see the effects discussed
in the previous section.

\section*{Acknowledgements}
This work was carried out under the PPARC Research Grant GR/L56039
and the Leverhulme Grant F618C.
The author is grateful to Peter Weiss for helpful discussions.

\newpage
{~~~~~~~~~~~~~~~~~~~~~~~~~~~~~~~}
\vskip 20 truemm
\vskip 20 truemm
\vskip 20 truemm
\vskip 20 truemm
\vskip 20 truemm

\end{document}